\documentstyle[12pt]{article}
\begin{document}
\pagestyle{plain}
\title{\bf Quantum mechanical corrections to the Alfv\'en waves}
\author{ Miroslav Pardy\\[5mm]
Department of Physical Electronics, \\and \\
The Laboratory of the Plasma Physics,\\
Masaryk University,\\
Kotl\'{a}\v{r}sk\'{a} 2, 611 37 Brno, Czech Republic\\
e-mail:pamir@physics.muni.cz}
\date{\today}
\maketitle

\vspace{30mm}

\begin{abstract}

The hydrodynamical model of quantum
mechanics based on the Schr\"{o}dinger equation is combined with the
magnetohydrodynamical term to form so called quantum
magnetohydrodynamic equation. It is shown that the quantum
correction to the Alfv\'en waves follows from this new equation. The
possible generalization is considered for the so called nonlinear 
Schr\"{o}dinger equation and for the situation where dissipation is
described by the Navier-Stokes equation.

\end{abstract}

\newpage

\section{Introduction}
\hspace{3ex}

The classical magnetohydrodynamics treats on the conductive liquids in
magnetic field. It describes the motion of the conductive liquids. On
the other hand, it describes also the ionized gases not
only in the laboratory conditions, but also in the cosmical space. It
involves such liquids as mercury, liquid sodium and so on. 

The hydrodynamical motion of the liquid induces the electric and
magnetic fields. At the same time the forces act on currents in the
magnetic fields and they influence substantially the motion of the liquid.
These currents change the original magnetic fields. It means that the
very complicated situation arises as a consequence of the interaction
between liquid medium and the magnetic and electric fields. The basic
classical problem is to formulate the magnetohydrodynamical equations
which describe the complex motion of such plasma medium.

It is well known that the basic magnetohydrodynamical equations have
been derived (Landau et al., 1982) and the question arises what is the
quantum description of the motion of the conductive liquid, or in
other words what are the quantum magnetohydrodynamical equations. Our
goal is to postulate such equation and to solve them for the simple
case of the Alfv\'en waves. We use the approximation where the quantum
description can be expressed as the quantum corrections to the
classical Alfv\'en waves. The similar known analogue is the
determination of the radiative corrections to the classical 
synchrotron radiation.

According to Madelung (1926) Bohm and Vigier (1954),
Wilhelm (1970), Rosen (1974, 1986) and
others, the original Schr\"{o}dinger equation can be transformed into the
hydrodynamical system of equations by using the so called Madelung ansatz:

$$\Psi={\sqrt n}\*e^{\frac{i}{\hbar}\*S},\eqno(1)$$
where $n$ is interpreted as the density of particles and $S$ is the classical
action for $\hbar\rightarrow 0$. The mass density is defined by relation
$\varrho=n\*m$ where $m$ is mass of~a~particle.

It is well known that after insertion of the relation (1) into the
original Schr\"{o}dinger equation

$$i \hbar \frac {\partial \Psi}{\partial t} =
 -  \frac {\hbar^2}{2m}\*\Delta \Psi + V\* \Psi,\eqno(2)$$
where $V$ is the potential energy, we get, after separating the real and
imaginary parts, the following system of equations:

$$\frac {\partial S}{\partial t} + \frac {1}{2m}\* (\nabla\* S)^2 + V =
\frac {\hbar^2}{2m} \frac {\Delta \sqrt{n}}{\sqrt{n}}\eqno(3)$$

$$\frac {\partial n}{\partial t} + {\rm div}(n\*{\bf v}) = 0 \eqno(4)$$
with

$${\bf v}=\frac {\nabla S}{m}. \eqno(5)$$

Equation (3) is the Hamilton-Jacobi equation with the additional term

$$V_q = - \frac {\hbar^2}{2m} \frac {\Delta \sqrt{n}}{\sqrt{n}}, \eqno(6)$$
which is called the quantum Bohm potential and equation (4) is the
continuity equation.

After application of operator $\nabla$ on eq. (3), it can
be cast into the Euler hydrodynamical equation of the form:

$$\frac{\partial {\bf v}}{\partial t}+({\bf v} \cdot \nabla)\* {\bf v}=
- \frac {1}{m}\*\nabla\* (V+V_q). \eqno(7)$$

It is evident that this equation is from the hydrodynamical point of view
incomplete as a consequence of the missing term
$-\varrho^{-1}\*\nabla\* p$
where $p$ is hydrodynamical pressure. We complete the
equation (7) by adding the pressure term and in such a way we get the
total Euler equation in the form:

$$m \left( \frac{\partial {\bf v}}{\partial t} + ({\bf v}\cdot\nabla)\*
{\bf v}\right)= - \nabla\*V_q - \frac{1}{n}\* \nabla p,\eqno(8)$$
where we have put $V = 0$.

In case of the magnetohydrodynamics, it is necessary to add the so
called magnetic term 

$$\frac{1}{4\pi n}{\bf H}\times {{\rm rot}\; \bf H}\eqno(9)$$
to the equation (8) in order to get

$$m \left( \frac{\partial {\bf v}}{\partial t} + ({\bf v}\cdot\nabla)\*
{\bf v}\right)= - \nabla\*V_q - \frac{1}{n}\* \nabla p -
\frac{1}{4\pi n}{\bf H}\times {{\rm rot}\; \bf H}.\eqno(10)$$

The last equation (10) with the equation (4) and equations 

$${\rm div}\;{\bf H} = 0\eqno(11)$$

$$\frac{\partial {\bf H}}{\partial t} = {\rm rot}\;({\bf v}\times 
{\bf H}) \eqno(12)$$ 
form the basic equations of the quantum magnetohydrodynamics.
 
Now, let us find the solution of the system (4), (10), (11) and (12) in
the form of waves. 

\section{Alfv\'en waves with quantum mechanical corrections}

The solution of the system is known in case that the quantum
mechanical potential is neglected. If we respect the q-potential, we
get with 

$${\bf H}= {\bf H}_{0} + {\bf h},\quad \varrho = \varrho_{0} + \varrho',\quad 
p = p_{0} + p' , \eqno(13)$$ 
the following equations   

$${\rm div}\;{\bf h} = 0, \quad \frac{\partial {\bf h}}{\partial t} = 
{\rm rot}({\bf v}\times {\bf h}),\quad 
\frac {\partial \varrho'}{\partial t} + {\rm div}(\varrho_{0}\*{\bf v}) = 0,
\eqno(14)$$

$$ \frac{\partial {\bf v}}{\partial t} = - \nabla\left( - \frac
  {\hbar^2}{2m^{2}} \frac {\Delta \sqrt{\varrho_{0} + \varrho'}}
{\sqrt{\varrho_{0}+\varrho'}}\right) - 
 \frac{1}{\varrho_{0}+\varrho'}\nabla p' -
\frac{1}{4\pi (\varrho_{0}+ \varrho')}{\bf H}_{0}\times {\rm rot}\;
  {\bf h} 
\eqno(15)$$

We see that in case of the incompressional fluid, the quantum
corrections are zero. So the incompressional fluid is not quantum
mechanical, but only classical.

Using the approximative relation

$$\nabla\left(\frac
  {\hbar^2}{2m^{2}} \frac {\Delta \sqrt{\varrho_{0}+\varrho'}}
{\sqrt{\varrho_{0}+\varrho'}}\right)\approx 
\frac{\hbar^2}{4m^{2}}\frac{1}{\varrho_{0}}{\rm grad}\Delta\varrho'\eqno(16)$$
and hydrodynamical relation

$$p' = u_{0}^{2}\varrho',\eqno(17)$$
where $u_{0}^{2}$ is the velocity of sound in the medium, we get 

$$ \frac{\partial {\bf v}}{\partial t} = 
\frac{\hbar^2}{4m^{2}}\frac{1}{\varrho_{0}}{\rm grad}\Delta\varrho' - 
\frac{ u_{0}^{2}}{\varrho_{0}}{\rm grad}\varrho' - 
\frac{1}{4\pi \varrho_{0}}({\bf
  H}_{0}\times {\rm rot}\;{\bf h}). \eqno(18)$$

The solution of equations (14) and (18) can be realized in the
form:

$${\bf v} = {\bf v}_{0}e^{i({\bf k}{\bf r} - \omega t)},\quad
{\bf h} = {\bf h}_{0}e^{i({\bf k}{\bf r} - \omega t)},\quad
\varrho' = \varrho'_{0}e^{i({\bf k}{\bf r} - \omega t)},\eqno(19)$$
where ${\bf k}$ is the wave vector and $\omega$ is a frequency of the 
wave. After insertion of eqs. (19) in equations (14) and (18),
we get 

$${\bf k}{\bf h} = 0 \eqno(20)$$

$$-\omega{\bf h} = {\bf k}\times({\bf v}\times{\bf h})\eqno(21)$$

$$\omega\varrho'= \varrho_{0}({\bf k}{\bf v})\eqno(22)$$

$$-\omega{\bf v} +\frac{u_{0}^{2}}{\varrho_{0}}\varrho'{\bf k}
=  - \frac{1}{4\pi \varrho_{0}}{\bf  H}_{0}\times ({\bf k}
\times{\bf h}) -
\frac{\hbar^2}{4m^{2}}\frac{\varrho'}{\varrho_{0}}
k^{2}\;{\bf k}.\eqno(23)$$

We see from (20) that perturbation of the magnetic field 
${\bf h}$ is
perpendicular  to the wave vector  ${\bf k}$. So, we can choose  
${\bf  k} \equiv (k_{x}, 0, 0)$  and  ${\bf H}_{0} \equiv
(H_{0x},H_{0y}, 0)$. Then, with $u=\omega/k$, we have the equations
following from eq. (21) and (23)(where in eq. (23) the term 
$\varrho'/\varrho_{0}$ was excluded using eq. (22)) (Landau et al., 1982).

$$uh_{z} = -v_{z}H_{0x}, \quad uv_{z} = 
-\frac{H_{0x}}{4\pi\varrho_{0}}h_{z},\eqno(24)$$ 

$$uh_{y} = v_{x}H_{0y} -  v_{y}H_{0x}, \quad uv_{y} =
-\frac{H_{0x}}{4\pi\varrho_{0}}h_{y},\eqno(25)$$ 

$$\left(u - \frac{U_{0}^{2}}{u}\right)v_{x} = 
\frac{H_{0y}}{4\pi\varrho_{0}}h_{y},\eqno(26)$$
where

$$U_{0}^{2} = u_{0}^{2} + \frac{\hbar^{2}}{4m^{2}}k^{2}.\eqno(27)$$

It follows from eq. (22) that 

$$\varrho' = \frac{\varrho_{0}}{u}v_{x}\eqno(28)$$

The necessary condition of solubility of eqs. (24) can be expressed
as

$$u = \frac{|H_{0x}|}{\sqrt{4\pi \varrho_{0}}} \equiv u_{A},\eqno(29)$$
where A is the first letter of Alfv\'en.

If we will suppose that $H_{0x}>0$ then we can remove the absolute value.
The velocity $v_{z}$ is also vibrating an it is easy to see that 

$$v_{z} = -\frac{h_{z}}{\sqrt{4\pi \varrho_{0}}}.\eqno(30)$$

Or, in general, 

$${\bf v} = - \frac{{\bf h}}{\sqrt{4\pi\varrho_{0}}}\eqno(31)$$
  
The equation (30) can be written in the more general form as 

$$\omega = \frac{1}{\sqrt{4\pi \varrho_{0}}} {\bf H_{0}}{\bf k},\eqno(32)$$
from which it follows the group velocity in the form 

$$\frac{\partial\omega}{\partial{\bf k}} = 
\frac{{\bf H_{0}}}{\sqrt{4\pi \varrho_{0}}}\eqno(33)$$
and it does not depend on the direction of $\bf k$. So the group
velocity of a wave is in the direction of the magnetic field $\bf
H_{0}$. These wave are so called Alfv\'en waves with the velocity
$u_{A}$. The analysis of the properties of the magnetohydrodynamical
waves is described in the Landau et al. textbook (Landau et al., 1982). 

Let us remark that from eqs. (25) and (26) it follows the so called
magnetosound waves as a result of the determinat of these equation
which we put zero. Or, (Landau et al., 1982)

$$U_{\sigma M}^{2} = \frac{1}{2}
\left\{\frac{H_{0}^{2}}{4\pi\varrho_{0}} + 
U_{0}^{2} \pm 
\left[\left(\frac{H_{0}^{2}}{4\pi\varrho_{0}} + U_{0}^{2}\right)^{2}
- \frac{H_{0x}^{2}}{\pi\varrho_{0}}U_{0}^{2}\right]^{1/2}\right\}.\eqno(34)$$

So, we get two waves involving the quantum corrections. The first wave
is quick wave with sign $+$  and the second wave is slow wave with
sign $-$.

\section{The nonlinear quantum magnetohydrodynamics equation}

\hspace{3ex}

In case of the nonlinear Schr\"{o}dinger equation with the logarithmic
nonlinearity the basic equation is of the form (Pardy, 2001;
Bialynicky-Birula et al., 1976):
  
$$i \hbar \frac {\partial \Psi}{\partial t} =
 -  \frac {\hbar^2}{2m}\*\Delta \Psi + V\* \Psi + b(\ln|\Psi|^2)\Psi,
\eqno(35)$$
where $b < 3\times10^{-15}eV$ (G\"{a}hler et al., 1981) is some constant. 

The quantum equation of MGH with the nonlinear term is then 

$$m \left( \frac{\partial {\bf v}}{\partial t} + ({\bf v}\cdot\nabla)\*
{\bf v}\right)=  \nabla\left(\frac
  {\hbar^2}{2m}\frac {\Delta \sqrt{n}}
{\sqrt{n}}\right) + b(\ln|n|^2) - \frac{1}{n}\* \nabla p -
\frac{1}{4\pi n}{\bf H}\times {\rm rot}\; {\bf H}. \eqno(36)$$
Or, 

$$ \left( \frac{\partial {\bf v}}{\partial t} + ({\bf v}\cdot\nabla)\*
{\bf v}\right)=  \nabla\left(\frac{\hbar^2}{2m^{2}} 
\frac {\Delta\sqrt{\varrho}}{\sqrt{\varrho}}\right) +
\frac{b}{m}(\ln\left|\frac{\varrho}{m}\right|^2) - 
\frac{1}{\varrho}\* \nabla p -
\frac{1}{4\pi \varrho}{\bf H}\times {\rm rot}\; {\bf H}. \eqno(37)$$

It is evident that to finding the quantum magnetohydrodynamical solutions
will be more complicated than of the linear QMGH problems. Let us
first remember the one-dimensional solutions of the one-dimensional
nonlinear Schr\"{o}dinger equation (Pardy, 2001).

Let be $c, ({\rm Im} \;c =0), v, k, \omega$ some parameters and 
let us insert function

$$\Psi (x,t)= c\* G (x-v\*t)\* e^{i\*k\*x-i\*\omega\*t}\eqno(38)$$
into the one-dimensional equation (35) with $V=0$. Putting the imaginary
part of the new equation to zero, we get

$$v= \frac{\hbar\*k}{m}\eqno(39)$$
and for function $G$ we get the following nonlinear equation (symbol
$'$
denotes
derivation with respect to $\xi= x-vt)$:

$$ G'' + A\*G + B(\ln{G})G = 0,\eqno(40)$$
where
$$A= \frac{2m}{\hbar}\*\omega - k^2 + \frac{2m}{\hbar^2}\*b \*\ln{c^2}
\eqno(41)$$

$$B= \frac{4mb}{\hbar^2}.\eqno(42)$$

After multiplication of eq. (40) by $G'$ we get:

$$\frac{1}{2}\*{\left[ G'^2 \right]}^{'} + \frac{A}{2}\*{\left[ G^2\right]}^{'}
+  B\* {\left[ \frac{G^2}{2} \ln{G} - \frac{G^2}{4} \right]}^{'} =0,\eqno(43)$$
or, after integration

$$G'^2=- AG^2 - BG^2 \ln{G} + \frac{B}{2}\* G^2 + const.\eqno(44)$$

If we choose the solution in such a way that $G(\infty)=0$ and
$G'(\infty)=0$, we get $const.=0$ and after elementary operations we get the
following differential equation to be solved:

$$\frac{dG}{G \sqrt{a-B\*\ln{G}}}= d\xi,\eqno(45)$$

where
$$a=\frac{B}{2} - A.\eqno(46)$$

Equation (45) can be solved by the elementary integration and the
result is

$$G= e^{\frac{a}{B}}\*e^{-\frac{B}{4}\*(\xi+d)^2},\eqno(47)$$
where $d$ is some constant.

The corresponding soliton-wave function is evidently in the one-dimensional
free particle case of the form:

$$\Psi(x,t)= 
c\*e^{\frac{a}{B}}\*e^{-\frac{B}{4}\*(x-vt+d)^2}\*e^{ikx-i\omega\*t}.
\eqno(48)$$

It is not necessary to change the standard probability interpretation of the
wave function. It means that the normalization condition in our one-dimensional
case is

$$\int_{-\infty}^{\infty}{\Psi^*\*\Psi\,dx} =1.\eqno(49)$$

Using the Gauss integral

$$\int_{0}^{\infty}{e^{-\lambda^2\*x^2}\,dx} =
\frac{\sqrt{\pi}}{2\lambda},\eqno(50)$$
we get with $\lambda= {\left(\frac{B}{2}\right)}^{\frac{1}{2}}$

$$c^2\*e^{\frac{2a}{B}}= {\left(\frac{B}{2\pi}\right)}^{\frac{1}{2}}\eqno(51)$$
and the density probability $\Psi^*\Psi = \delta_m(\xi) $ is of the form
(with $d=0$):

$$\delta_m(\xi)= \sqrt{\frac{m\alpha}{\pi}}\* e^{-\alpha m \xi^2}
\hspace{5mm};\hspace{5mm}\alpha= \frac{2b}{\hbar^2}.\eqno(52)$$

It may be easy to see that $\delta_m(\xi)$ is the delta-generating function and
for $m \rightarrow \infty$ is just the Dirac $\delta$-function.

It means that the motion of a particle with sufficiently big mass $m$ is
strongly localized and in other words it means that the motion of this particle
is the classical one. Such behavior of a particle cannot be obtained in the
standard quantum mechanics because the plane wave
$\exp[ikx-i\omega\*t]$
corresponds to the free particle with no possibility of localization
for $m \rightarrow \infty$.

Let us still remark that it is possible to show that 
coefficient $c^2$ is real and positive number (Pardy. 2001).
 
We frequently read in the physical texts on the quantum mechanics that 
the classical limit of quantum mechanics is obtained only by the so called
WKB method. However, the limit is only formal because in this case
the probabilistic form of the solution is conserved while classical
mechanics is strongly
deterministic. In other words, statistical description of quantum 
mechanics is in no case reduced to the strong determinism of 
classical mechanics of one-particle system. So, only nonlinear 
quantum mechanics of the above form gives the correct classical 
limit expressed by the delta-function. More information on the
problems which are solved by the nonlinear  Schr\"{o}dinger  equation
involving the collapse of the wave function and the Schr\"{o}dinger
cat paradox is described in author's articles (Pardy, 2001, 1994). The
extended version of the nonlinear  quantum world is described
in the preprint of Castro (2002).

\vspace{5mm}

\section{Discussion}
\hspace{3ex}

We have seen that the quantum corrections to the Alfv\'en waves
follows from the hydrodynamical formulation of quantum mechanics involving
the magnetic interaction and pressure. The  Alfv\'en
waves involve the Planck constant.

The theory  can be generalized  to the situation with dissipation.
The dissipative terms can
be evidently inserted in the equation (10) in order to get quantum
Navier-Stokes magneto-hydrodynamical equation:

$$m\left(\frac{\partial {\bf v}}{\partial t} + ({\bf v}\cdot \nabla){\bf v}
\right) = -  \nabla V_{q} - \frac{1}{n} \nabla p \quad + $$

$$\frac{\eta}{n}\Delta{\bf v} + \frac{1}{n}\left(\xi + \frac{\eta}{3}\right)
{\rm grad}\;{\rm div}\;{\bf v}
+ \frac{1}{4\pi n}({\rm rot}{\bf H}\times{\bf H}), 
\eqno(53)$$
where $\eta$ and $\xi$ are some classical constant which express the
dissipation. 

The Alfv\'en waves follow from the equation (53) in the
approximative form in such a way they will involve the dissipation and
the quantum corrections. We know that the Navier-Stokes equation are
used in the determination of the turbulence. If we apply such methods
to our system with the quantum corrections, we evidently get the so
called quantum turbulence. To our knowledge such problem was not
solved and it is not involved in the monographs on turbulence and
monographs dealing with the theory of catastrophes. It is not excluded
that the quantum turbulence plays the substantial role in the natural 
terrestrial catastrophes.

\vspace{15mm}
\noindent
{\bf References}

\vspace{5mm}

\noindent
Bialynicky-Birula I. and  Mycielski, J. Nonlinear wave mechanics,
Ann. Phys. (N.Y.) 100 (1976) 62.\\[2mm]
Bohm D. and  Vigier, J.  Model of the causal interpretation of
quantum theory in terms of a fluid with irregular fluctuations,
Phys. Rev. 96 No. 1 (1954) 208.\\[2mm]
Castro, C. J. Mahecha  and Rodr\'iguez B.,
Nonlinear QM as a fractal Brownian motion with complex diffusion
constant, quant-ph/0202026.\\[2mm]
G\"{a}hler, R.  Klein  A. G. and  Zeilinger, A. Neutron optical
test of nonlinear wave mechanics,  Phys. Rev. A 23 No. 4 (1981) 1611.\\[2mm]
Landau, L. D. and  Lifshitz, E. M. Electrodynamics of the continuous
media, Moscow, Nauka (1982). (in Russian).\\[2mm]
Madelung, E. Quantentheorie in hydrodynamischer Form,
Z. Physik 40 (1926) 322.\\[2mm]
Pardy, M. To the nonlinear quantum mechanics,
quant-ph/0111105.\\[2mm]
Pardy, M. Possible Tests of Nonlinear Quantum Mechanics,
in: Waves and Particles in Light and Matter, Ed. by Alwyn van der Merwe and
Garuccio, A. Plenum Press New York (1994).\\[2mm]
Rosen, N. A classical picture of quantum mechanics,
Nuovo Cimento 19 B  No. 1 (1974) 90.\\[2mm]
Rosen, N. Quantum particles and classical particles,
Foundation of Physics 16 No. 8 (1986) 687.\\[2mm]
Wilhelm, E. Hydrodynamic model of quantum mechanics,
Phys. Rev. D 1 No. 8 (1970) 2278.

\end{document}